\documentclass[10pt]{iopart}
\usepackage{graphicx}
\begin{document}

\title{Quantum fluctuations in the mazer}

\author{Jonas Larson}

\address{NORDITA, 10691 Stockholm, Sweden}

\ead{jolarson@kth.se}

\begin{abstract}

Quantum fluctuations in the mazer are considered, arising either from
the atomic motion or from the quantized intracavity field.
Analytical results, for both the meza and the hyperbolic secant mode
profile, predict for example an attenuation of tunneling resonances
due to such fluctuations. The case of a Gaussian mode profile is
studied numerically using a wave packet propagation approach. The
method automatically takes into account fluctuations in the
atomic motion and the dynamics is especially considered at or
adjacent to a tunnel resonance. We find that the system evolution is
greatly sensitive to the atom-field detuning, bringing about a
discussion about the concept of adiabaticity in this model. Further,
a novel collapse-revival phenomena is demonstrated, originating from
the quantum fluctuations in the atomic motion rather from field
fluctuations as is normally the case.

\end{abstract}
%\maketitle

\section{Introduction}\label{sec1}

In recent years, the field of experimental cavity quantum
electrodynamics (QED) has expanded beyond the setup of single
thermal two-level atoms traversing the cavity one by one
\cite{haroche1,cqed1,haroche2}. Cavity QED now includes the
realization of solid state q-dots coupled to cavity fields
\cite{qdot}, trapped ion-cavity systems \cite{ioncavity},
Bose-Einsten condensates coherently interacting with single cavity
modes \cite{bec} and ultracold atomic many-body systems inside
cavities; {\it e.g.}, bistability behavior \cite{sk}, cavity cooling
\cite{cool} and self-organization of atoms in regular patterns
\cite{so}. Reaching these new regimes of cavity QED has been
possible due to the progress in the cooling, trapping and coherent
control of atoms \cite{meystre}. Actually, atomic velocities of the order of 1 cm/s or even slower are achievable \cite{atvel}. This opens up the possibility to
experimentally demonastrate the {\it mazer} (Microwave Amplification via
$z$-motion-induced Emission of Radiation), where ultracold, rather
than thermal, two-level atoms are let to pass the cavity
\cite{mazer1,mazer2}. 

In the ultracold regime, the kinetic energy of the atom is of the
order of the atom-field interaction energy and must be treated
quantum mechanically. Early on, it was predicted that atoms can be
back scattered by the high-$Q$ cavity vacuum induced potential
\cite{mazer1} or even trapped by the intracavity field
\cite{harochetrap}, which indeed was later experimentally obtained
for single photon fields \cite{cavitytrap}. Later, in a series of
papers by the late Walther and coworkers, it was shown that in this
new regime the emission probability for an excited atom traversing
the cavity is greatly affected by the quantized atomic motion
\cite{mazer2}. For certain resonance conditions, where the atomic de
Broglie wavelength match resonantly with the cavity length, the
atom tunneling and emission probabilities are greatly increased.
Such resonances lead to new phenomena of the field statistics in the
micromaser pumped by excited atoms. Following \cite{mazer2}, a set
of papers where published on the mazer. Most of which consider
analytically solvable mazer models, namely the atom-field detuning
vanishes and the cavity mode profile has the shape of a meza function or a
hyperbolic secant \cite{mazer3,mazer4,mazergauss,mazerpoptrap,tunneltime}. These all assume the zero detuning situation where the mazer problem can be decoupled into two separate 1-D scattering problems,
while for a non-zero detuning the dynamics is indeed more complex and the
full coupled system must be considered. This was the subject of \cite{mazerdet}, where the problem of a non-zero detuning in the
case of a meza function coupling was solved analytically (see also \cite{muga2}). Further, the situation of a sinusoidal mode profile with one or a few
wavelengths and a Gaussian shape have been analyzed by means of approximate or numerical methods in \cite{sin} and \cite{mazergauss} respectively. The limit in which the mode profile is "delta"-shaped has as well been discussed \cite{mazerdelta,jonas6}. The mazer problem for a sinusoidal coupling with several wavelengths
was investigated in \cite{jonas5} and the model was found to exhibit
very different characteristics due to the quasi-periodicity, see
also \cite{sin2}. Extensions of the regular mazer setup have been
considered in numerous works, {\it e.g.} a driven system
\cite{drivemazer}, multi-level atoms and multi-photon processes
\cite{multilevel}, multi-cavity setups \cite{mazer3,muga2,multicav}, Kerr
and gravity effects on the mazer action \cite{kerr,grav}. Also
semi-classical approaches have been studied, where for example the
atomic motion is replaced by an effective time-dependence
\cite{jonas7,jonas4}.

None of the above references take into account fluctuations in
the motion of the atom. In the standard atom cavity QED experiment
in which the atom traverses the cavity, its position is fairly well localized
in space, typically for a microwave setup within the order of 1 mm
(where the mode waist $L$ is about 5-6 mm) \cite{haroche1}. Thus, in
real experiments is the state of the atom given by a well localized
wave packet rather than a traveling wave. Naturally, an uncertainty
in position results in fluctuations in momentum which, on the other
hand, affects the particle scattering. In this paper we address these
issues both by generalizing known analytical models but also via
numerical analysis of a Gaussian mode profile. The Gaussian coupling
approximates the mode profile of a Fabry-Perot cavity which is
usually used in micromaser experiments \cite{haroche1,cqed1}. A
crucial feature of the Gaussian, as well for the hyperbolic secant,
coupling is that they are smooth with a well defined analytical
limit, contrary to the meza function. This give rise to qualitative
different results compared with the meza case. Here we focus on
these consequences and discuss the concept of adiabaticity in the
mazer. One essential parameter for adiabaticity is the
atom-field detuning and how it affects the dynamics is investigated.
It is found that it plays a central role for the evolution. So far, non-zero detuning has only been considered for meza-mode couplings, which do not possess an adiabatic limit due to their discontinuities. Quantum fluctuations lead in
general to a smearing out of physical observables, but we demonstrate that it can also render new collapse-revival structures.

The general mazer concept and the model system are introduced in the section \ref{sec2}. The various reflection and transmission
coefficients, as well as appropriate bases, are given. Our
analytical results are presented in section \ref{sec3}, where
quantum fluctuations in both the meza and hyperbolic secant cases are
considered. The following section \ref{sec4} is devoted to numerical
research of the Gaussian mode profile for zero and non-zero
detunings. We apply a split-operator approach that invitably takes into account
fluctuations in the motion of the atom. The effect of the
detuning parameter is studied as well as the system dynamics during
the scattering process. In the final section \ref{sec5} we give a
summery and discuss experimental prospects.

\section{The model system}\label{sec2}
We consider a two-level atom with internal states $|g\rangle$ and
$|e\rangle$ and center-of-mass (c.o.m.) momentum and position
$\hat{p}$ and $\hat{z}$, which dipole interacts with a single
quantized mode of a cavity. The effective Hamiltonian in the
rotating wave approximation reads
\begin{equation}\label{ham1}
\tilde{H}=\frac{\hat{p}^2}{2m}+\hbar\omega\left(\hat{a}^\dagger\hat{a}+\frac{1}{2}\right)+\frac{\hbar\Omega}{2}\hat{\sigma}_z+\hbar\lambda(\hat{z})\left(\hat{a}^\dagger\hat{\sigma}^-+\hat{\sigma}^+\hat{a}\right),
\end{equation}
where $m$ is the mass of the atom, $\omega$ the field frequency,
$\Omega$ is the atomic transition frequency and $\lambda(\hat{z})$
is the effective position dependent atom-field coupling. Here
$\hat{a}$ and $\hat{a}^\dagger$ are the annihilation and creation
operators for the field and the $\sigma$-operators are the regular
Pauli matrices, $\hat{\sigma}_z=|e\rangle\langle e|-|g\rangle\langle
g|$, $\hat{\sigma}^-=|g\rangle\langle e|$ and
$\hat{\sigma}^+=|e\rangle\langle g|$. The number of excitations is
conserved and it may be used to define an interaction picture rotating with
$\omega$. This introduces the atom-field detuning
$\Delta=\Omega-\omega$ and the transformed Hamiltonian becomes (we let $\hbar=m=1$)
\begin{equation}\label{ham2}
H=\frac{\hat{p}^2}{2}+\frac{\Delta}{2}\hat{\sigma}_z+ \lambda(\hat{z})\left(\hat{a}^\dagger\hat{\sigma}^-+\hat{\sigma}^+\hat{a}\right).
\end{equation}
Due to this symmetry can the Hamiltonian be written on
$2\times2$ block diagonal form within its internal states. For a
given number of excitations, corresponding to one of the
$2\times2$-blocks, it is often convenient to turn to a {\it dressed
representation} by transforming $H$ with \cite{jonas2}
\begin{equation}\label{unitrans}
U=\left[\begin{array}{cc} \cos(\theta) & \sin(\theta)\\
-\sin(\theta) & \cos(\theta)
\end{array}\right],
\end{equation}
where $\tan(2\theta)=2\lambda(\hat{z})\sqrt{n}/\Delta$. This defines
the {\it dressed internal states}
\begin{equation}\label{eigstate}
\begin{array}{l}
|\gamma_n^+\rangle=\cos(\theta)|n,e\rangle+\sin(\theta)|n+1,g\rangle,\\ \\
|\gamma_n^-\rangle=-\sin(\theta)|n,e\rangle+\cos(\theta)|n+1,g\rangle.
\end{array}
\end{equation}
Here, $|n,e\rangle$ and $|n+1,g\rangle$ are the states with $n$
photons and the atom excited or $n+1$ photons and the atom in its
ground state respectively. These states are referred to as {\it bare
internal states}. The bare state $|0,g\rangle$ is decoupled from any
other states, which, however is not true beyond the rotating wave
approximation. We further have that $U$ diagonalizes the last two
terms in the Hamiltonian (\ref{ham2}) whose eigenvalues read
\begin{equation}
V_n^\pm(z)=\pm\sqrt{\left(\frac{\Delta}{2}\right)^2+\lambda^2(z)(n+1)},\hspace{1cm}n=0,1,2,...\,.
\end{equation}
We call these {\it adiabatic potentials} whenever $\Delta\neq0$. Since
the operator $U$ is $\hat{z}$-dependent it will not commute with the
first term of $H$, causing non-diagonal terms which couples the two
dressed internal states \cite{jonas2}. However, in the limiting
cases $\Delta\rightarrow\pm\infty,\,0$ one has that these {\it
non-adiabatic couplings} vanish. Thus, for $\Delta=0$ we may
decouple the two equations and we note that in this special case
$\cos(\theta)=\sin(\theta)=1/\sqrt{2}$. We distinguish this limit
with the ones of large detuning ({\it adiabatic limit}) and call it
{\it diabatic limit} \cite{jonas3}.

Due to the coupled two-level structure, the dynamics may become
rather complicated. An atom entering the interaction region will in general
feel both an attractive and a repulsive potential. Despite the
coupled dynamics, transmission and reflection coefficients for the two internal states can be well defined. In particular, for the
dressed basis we call these $\rho_n^\pm$ and $\tau_n^\pm$. It
follows that the coefficients in the bare basis are related to the
dressed ones via \cite{mazer2}
\begin{equation}\label{coef1}
\begin{array}{c}
\displaystyle{R_{en}=\frac{1}{2}\left(\rho_n^++\rho_n^-\right)},\hspace{1.2cm}\displaystyle{R_{gn}=\frac{1}{2}\left(\rho_n^+-\rho_n^-\right)},\\ \\
\displaystyle{T_{en}=\frac{1}{2}\left(\tau_n^++\tau_n^-\right)},\hspace{1.2cm}\displaystyle{T_{gn}=\frac{1}{2}\left(\tau_n^+-\tau_n^-\right)}.
\end{array}
\end{equation}
The $R$-coefficients squared give the probability for reflection of
the atom in state $e$ or $g$ respectively and correspondingly $T$
gives the transmission. Thus, we have the transmission probability,
given an excitation $n$,
\begin{equation}\label{probtrans1}
P_{trans}^n=|T_{en}|^2+|T_{gn}|^2
\end{equation}
and similarly for the reflection. It should be pointed out that the
mazer scattering process is different from the regular 1-D ones
often presented in textbooks \cite{scatbook}. This comes about
because of the coupled dynamics between the internal states, and in
particular, in some limiting cases can the particle be said to be
scattered simultaneously to an attractive and repulsive potential.
This, in fact, leads to new phenomena \cite{mazer3}, not encountered in the standard situations. Another
property of interest is the {\it von
Neumann entropy} which gives a measure of entanglement shared
between the field and the atom. Defining the reduced density
operator for the field as $\rho_f=\mathrm{Tr}_{at}\big[\rho\big]$,
where $\rho$ is the full system density operator and the trace is
over atomic degrees of freedom, one has the von Neumann entropy
\begin{equation}
S_n=-\mathrm{Tr}\big[\rho_f\log(\rho_f)\big].
\end{equation}
The field and atomic entropies are identical for initial pure states
\cite{araki}. In terms of the coefficients (\ref{coef1}) it follows that
\begin{equation}\label{entropy1}
\begin{array}{lll}
S_n & = & -\left(|R_{en}|^2+|T_{en}|^2\right)\log\left[\left(|R_{en}|^2+|T_{en}|^2\right)\right]\\ \\
& & -\left(|R_{gn}|^2+|T_{gn}|^2\right)\log\left[\left(|R_{gn}|^2+|T_{gn}|^2\right)\right].
\end{array}
\end{equation}

Let us choose the initial state to be the uncorrelated state with
the field in vacuum, the atom in its internal excited state and with
some spatial wave function $\psi(z)$,
\begin{equation}
\langle z|\Psi(t=0)\rangle=\psi_i(z)|0,e\rangle.
\end{equation}
Using the position-momentum relation
\begin{equation}
\psi(z)=\int\,dk\,\tilde{\psi}(k)\mathrm{e}^{ikz}
\end{equation}
we can write down the wave function for large times, when the atom
is far from the interaction regime, as \cite{mazer2}
\begin{equation}
\begin{array}{lll}
\displaystyle{\langle z|\Psi(t)\rangle} & = & \displaystyle{\int\,dk\tilde{\psi}(k)\mathrm{e}^{-i(k^2/2)t}\left\{\left[R_{e0}(k)\mathrm{e}^{-ikz}+T_{g0}(k)\mathrm{e}^{ikz}\right]|0e\rangle\right.} \\ \\
& & +\displaystyle{\left.\left[R_{g0}(k)\mathrm{e}^{-ikz}+T_{g0}(k)\mathrm{e}^{ikz}\right]|1,g\rangle\right\}.}
\end{array}
\end{equation}
Taken into account for fluctuations in momentum we define the {\it
total} reflection and transmission coefficients
\begin{equation}\label{coef2}
\begin{array}{l}
\displaystyle{\left|\tilde{R}_{i0}\right|^2=\int\,dk\,\left|\tilde{\psi}(k)R_{i0}(k)\right|^2,\hspace{1.2cm}i=e,\,g},\\ \\
\displaystyle{\left|\tilde{T}_{i0}\right|^2=\int\,dk\,\left|\tilde{\psi}(k)T_{i0}(k)\right|^2,\hspace{1.2cm}i=e,\,g}.
\end{array}
\end{equation}
With these, the transmission probability or the von Neumann entropy
are readily modified. Noteworthy is that the coefficients (\ref{coef2}) is
as well easily generalized for general initial cavity fields,
$|\phi\rangle=\sum_nc_n|n\rangle$ by weighting the corresponding
coefficients by $|c_n|^2$ and summing over $n$.

\begin{figure}[ht]
\begin{center}
\includegraphics[width=10cm]{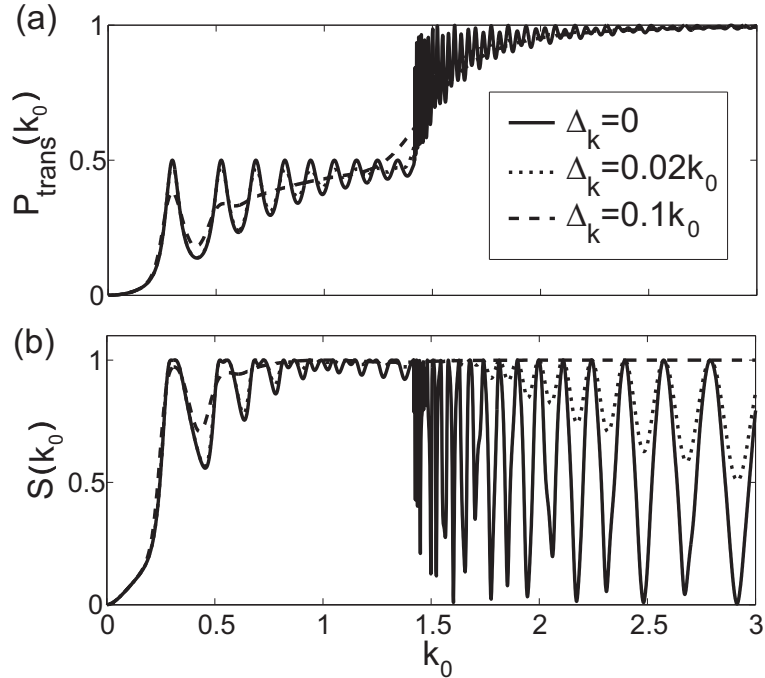}
\caption{\label{fig1} The transmission probability $P_{trans}$ and
the field entropy $S$ (for the meza function) as function of atomic
c.o.m. momentum $k_0$. Solid lines correspond to the results
(\ref{probtrans1}) and (\ref{entropy1}), while dashed and dotted
lines take into account for momentum fluctuations. Here
$\lambda_0=1$ and $L=50$.  }
\end{center}
\end{figure}

\section{Analytical consideration}\label{sec3}

Some of the outcomes from quantum fluctuations, both in the
atomic c.o.m. motional state and in the cavity photon field, are
demonstrated in this section using analytically solvable models. The sequential section continues the analysis by a numerical
study of a more physically realistic situation. In both models
considered here is the detuning $\Delta$ is assumed zero. In other words, the
internal structure can be separated into two decoupled equations.
The atomic motional wave function is taken to be a Gaussian,
centered at $k_0$ and with width $\Delta_k$,
\begin{equation}\label{initialwp}
\tilde{\psi}(k)=\left(\frac{1}{\pi\Delta_k^2}\right)^{1/4}\mathrm{e}^{-\frac{(k-k_0)^2}{4\Delta_k^2}}.
\end{equation}
If not mentioned, the cavity field is initially in the vacuum.

\subsection{Meza function}
The simplest solvable model, and also most extensively studied,
assumes a constant atom-field coupling within
and zero outside the cavity region
\cite{mazer1,mazer2,mazer3,mazer4}
\begin{equation}
\lambda(z)=\left\{\begin{array}{ll}
\lambda_0\hspace{0.8cm} & \mathrm{for}\,\,0<z<l\\
0\hspace{1.2cm} & \mathrm{elsewhere}.
\end{array}\right.
\end{equation}
The dressed transmission and reflection coefficients read
\cite{mazer2}
\begin{equation}
\begin{array}{c}
\rho_n^\pm=i\Delta_n^\pm\sin(k_n^\pm L)\tau_n^\pm,\\ \\
\tau_n^\pm=\exp(-ikL)\big[\cos(k_n^\pm L)-i\Sigma_n^\pm\sin(k_n^\pm L)\big]^{-1},
\end{array}
\end{equation}
where
\begin{equation}
\begin{array}{c}
k_n^\pm=\sqrt{k^2\mp\kappa_n^2},\\ \\
\Delta_n^\pm=\frac{1}{2}\left(\frac{k_n^\pm}{k}-\frac{k}{k_n^\pm}\right),\\ \\
\Sigma_n^\pm=\frac{1}{2}\left(\frac{k_n^\pm}{k}+\frac{k}{k_n^\pm}\right),\\ \\
\kappa=\sqrt{2\lambda_0},\\ \\
\kappa_n=\kappa\sqrt[4]{n+1}.
\end{array}
\end{equation}
The discontinuity of the atom-field coupling, once the problem is presented in the dressed basis, will render delta-like
non-adiabatic coupling terms \cite{jonas2}. This singularity
naturally affects the system properties and in particular are such effects believed to be absent in realistic experimental models to be discussed in the next Section.

\begin{figure}[h]
\begin{center}
\includegraphics[width=10cm]{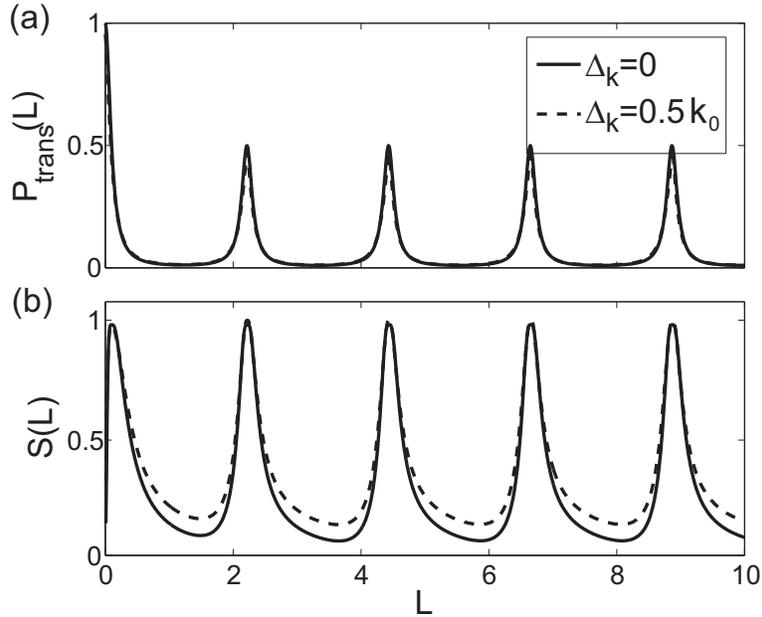}
\end{center}
\caption{\label{fig2} The transmission probability $P_{trans}$ and
the field entropy $S$ (for the meza function) as function of mode
length $L$. As in Figure \ref{fig1}, solid lines display the results
(\ref{probtrans1}) and (\ref{entropy1}), while the others include
momentum fluctuations. Here $\lambda_0=1$ and $k_0=\lambda_0/10$.  }
\end{figure}

The results for the transmission probability and the von Neumann
entropy are presented in Figures \ref{fig1} and \ref{fig2}. The
first figure shows how the quantities depend on the atomic c.o.m.
momentum $k_0$. The solid lines give the zero fluctuation case
(\ref{probtrans1}) and (\ref{entropy1}), the dotted lines are
calculated with a momentum fluctuation of 2 $\%$ of the c.o.m.
momentum, $\Delta_k=0.02k_0$, and for the dashed line
$\Delta_k=0.1k_0$. It is clear how the quantum fluctuations smear
out any variations of the two quantities, including the tunneling
resonances in the mazer regime $0<k_0<\sqrt{2}$. The atom-field
coupling is here $\lambda_0=1$ such that for momentum
$k_0>\kappa_0=\sqrt{2\lambda_0}$ is the atomic kinetic energy larger
than the potential barrier and the atomic transmission is abruptly
increased beyond $\kappa_0$. In this Rabi regime one notes how the
atom-field {\it Rabi oscillations} begin to form. The period of Rabi
oscillations decreases faster than the width $\Delta_k$ increases,
resulting in that these oscillations are captured in the dotted line
($\Delta_k=0.02k_0$) for large $k$. The atom-field entanglement is
maximized when the transition probability
reaches $1/2$ in the mazer regime. In the Rabi regime, the entanglement follows the swapping of the excitation between the field and the atom. In Figure \ref{fig2}, we show the same quantities when instead the coupling
length $L$ is varied. In this case, the results are insensitive to small momentum fluctuations and we therefore only show the results for
$\Delta_k=0.5k_0$. The resonances envisaged in the plots occur for $L=m\pi/\kappa_{n=0}$. Note that the fluctuations tend to increase the quantum correlations between the atom and the field.

\begin{figure}[h]
\begin{center}
\includegraphics[width=10cm]{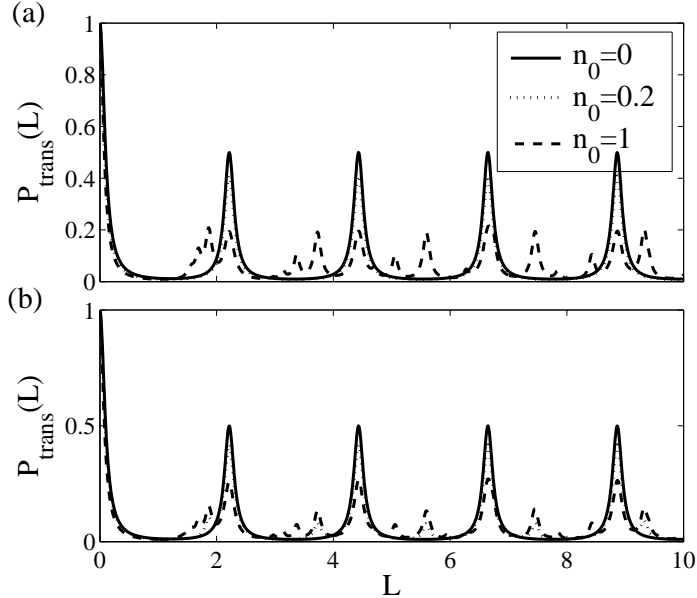}
\end{center}
\caption{\label{fig3} The transmission probability $P_{trans}$ (for
the meza function) as function of mode length $L$ for an initial
coherent cavity field (a) and an initial thermal cavity field (b).
The average number of photons is $n_0=0$ (solid line), $n_0=0.2$ or
$n_0=1$ (dashed line) and the other parameters are as in Figure
\ref{fig2}. }
\end{figure}

How an uncertainty in the field statistics affects the transmission
probability is depicted in Figure \ref{fig3}. The upper plot (a)
shows the transmission probability for an initial coherent field
state
\begin{equation}
|c_n|^2=\frac{n_0^n\mathrm{e}^{-n_0}}{n!}
\end{equation}
with mean photon number $n_0=0$, $n_0=0.2$ and $n_0=1$. Figure 3 (b) gives
the same but for an initial thermal field state
\begin{equation}
|c_n|^2=\frac{n_0^n}{(n_0+1)^{n+1}}.
\end{equation}
It is found that the tunneling resonances are very sensitive to
fluctuations in the cavity field. However, in the microwave regime of cavity
QED, a thermal field with $n_0=1$ ($n_0=0.2$) correspond to a
temperature of $0.5$ ($0.2$) degrees Kelvin, which is way above
current effective experimental temperatures \cite{haroche1}. The
secondary peaks showing up for non-zero $n_0$ are the tunneling
resonances arising from other photon quantum numbers $n$ differing
from $n=0$.

\subsection{Hyperbolic secant function}

\begin{figure}[h]
\begin{center}
\includegraphics[width=10cm]{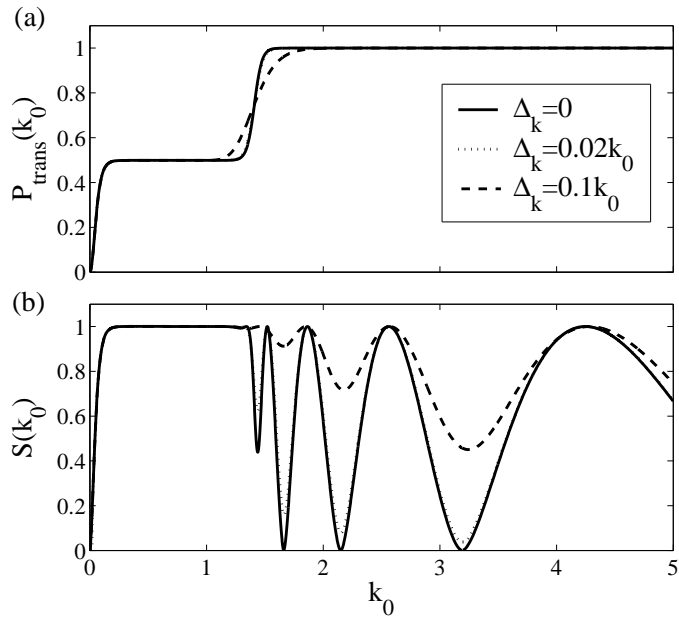}
\end{center}
\caption{\label{fig4} The transmission probability $P_{trans}$ and
the field entropy $S$ (for the hyperbolic secant function) as
function of the atomic c.o.m. momentum $k_0$. Contrary to the meza
function, the transition probability equals $1/2$ and the the atom-field becomes maximally entangled for most of the momenta $k_0$ in the mazer regime $k_0<\lambda_0\sqrt{2}$. Here, as in Figure
\ref{fig1}, $\lambda_0=1$ and $L=5$.  }
\end{figure}

To simulate a more realistic situation we consider a smooth
atom-field coupling
\begin{equation}
\lambda(z)=\lambda_0\,\mathrm{sech}^2\left(\frac{z}{L}\right),
\end{equation}
where $\lambda_0$ is the effective atom-field coupling and $L$
determines the mode waist length. The scattering problem analytically solvable in the special case of zero detuning $\Delta=0$, with dressed reflection and transmission coefficients \cite{mazer2}
\begin{equation}
\begin{array}{c}
\displaystyle{\rho_n^\pm(k)=\frac{\Gamma(ipL)\Gamma(1-ikL)}{\Gamma(1/2+i\xi_n^\pm)\Gamma(1/2-i\xi_n^\pm)}\tau_n^\pm,}\\
\\
\displaystyle{\tau_n^\pm(k)=\frac{\Gamma[1/2-i(ikL+\xi_n^\pm)]\Gamma[1/2-i(kL-\xi_n^\pm)]}{\Gamma(-ikL)\Gamma(1-ikL)},}
\end{array}
\end{equation}
where
\begin{equation}
\xi_n^\pm=\sqrt{\pm2\lambda_0L^2\sqrt{n+1}-1/4}.
\end{equation}
In terms of $\kappa_n$ (as defined above), the tunneling resonances
are given by $\kappa_nL=\sqrt{m(m+1)}$ for $m=1,2,3,...$
\cite{mazer2}. The transmission probability $P_{trans}^n(k_0)$ and
the von Neumann entropy $S(k_0)$ are displayed in Figure~\ref{fig4} as
function of $k_0$. The solid line gives the quantities for zero
momentum fluctuations, while the dotted and dashed curves take into
account for an uncertainty in the momentum distribution. It is seen that the transmission probability is constant in the mazer regime for this
smooth coupling. In the Rabi regime is the transmission
probability unity throughout composite to the meza function, which
is an effect of the smoothness of the coupling. From the entanglement
presented in plot (b) is it clear that in this regime the
system Rabi oscillates. Comparing Figures \ref{fig1} and \ref{fig4},
the effect of the non-analyticity of the meza function coupling is
visible from the irregular behaviour of Figure \ref{fig1}. Again,
the momentum fluctuations tend to blur the variations of the
curves.

\begin{figure}[h]
\begin{center}
\includegraphics[width=10cm]{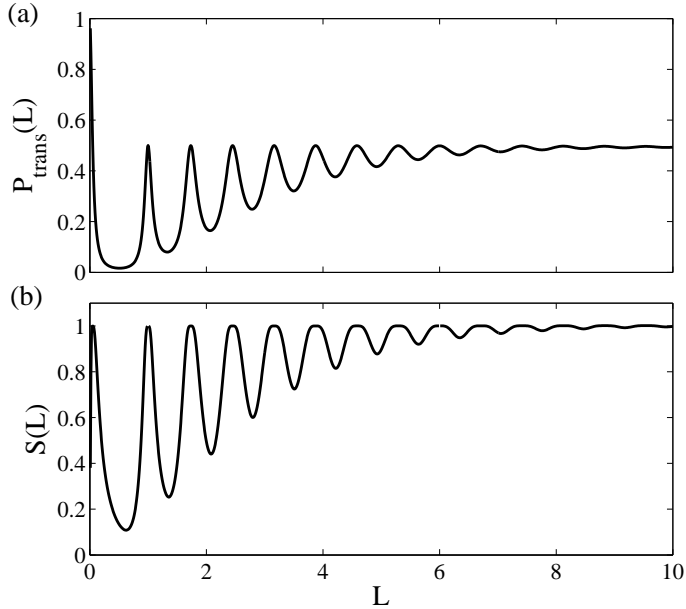}
\end{center}
\caption{\label{fig5} The transmission probability $P_{trans}$ and
the field entropy $S$ (for the hyperbolic secant function) as
function of mode waist $L$. Contrary to the meza function, the
tunneling resonances vanish for large $L$ where the transition
probability approaches $1/2$ and the the atom-field system becomes
maximally entangled, $S=1$. Here, as in Figure \ref{fig2}, are
$\lambda_0=1$ and $k_0=\lambda_0/10$.  }
\end{figure}

The dependence of the cavity mode waist $L$ is depicted in Figure
\ref{fig5}. As $L$ increases the transmission probability
$P_{trans}(L)$ and von Neumann entropy $S(L)$ approaches 1/2 and 1
respectively, which therefore deviate considerably from the results
for the meza function presented in Figure \ref{fig2}. It is known
that for large $L$, the adiabaticity is in general increased
\cite{jonas2}. Here, however, do we consider the case of $\Delta=0$
and we thus are in the strict diabatic limit rather than the
adiabatic one.

\section{Numerical consideration}\label{sec4}

\begin{figure}[h]
\begin{center}
\includegraphics[width=10cm]{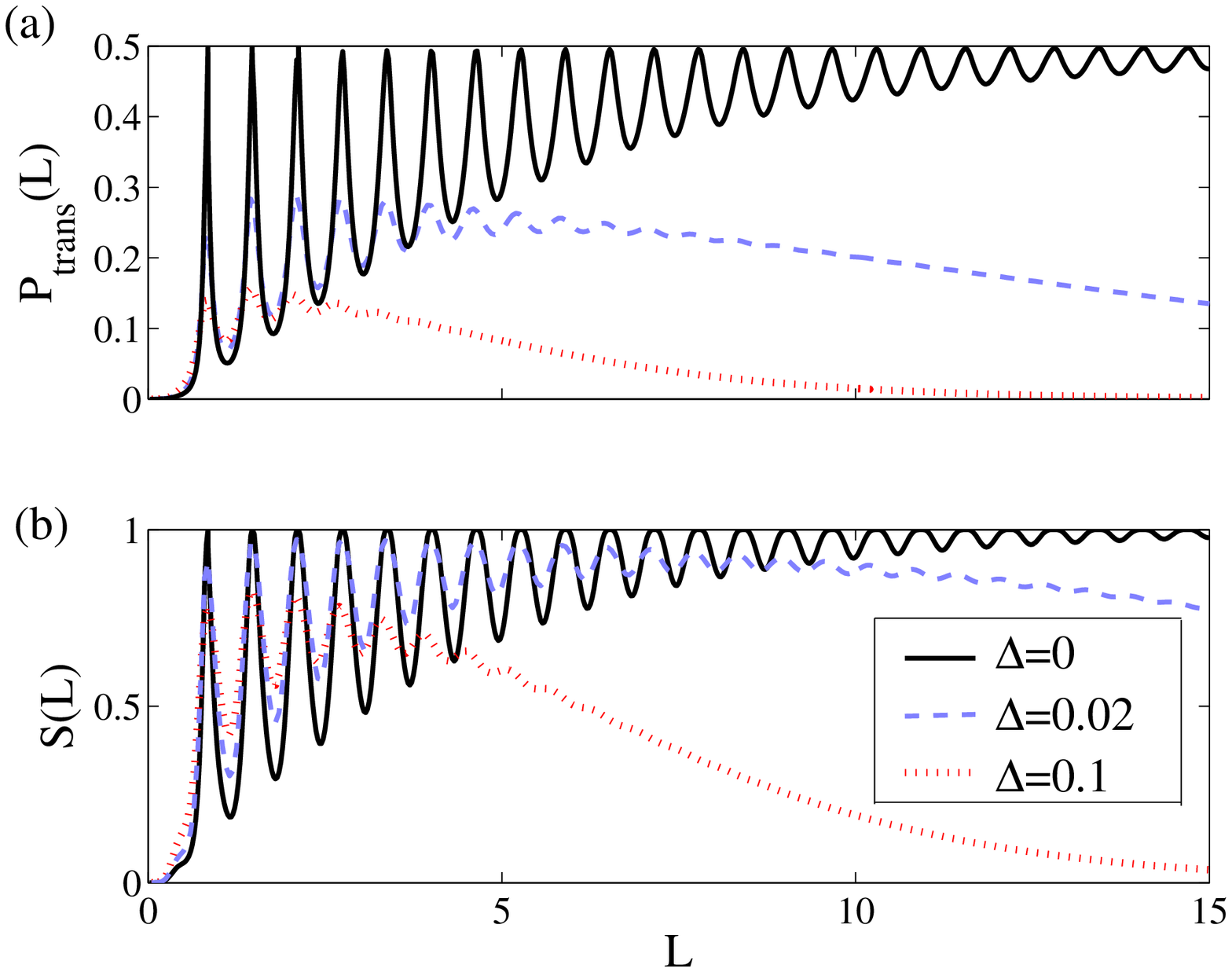}
\end{center}
\caption{\label{fig6} (Colour online) The transmission probability
$P_{trans}$ and the von Neumann entropy $S$ (for the Gaussian mode
profile) as function of mode waist $L$ and for three different
detunings; $\Delta=0$ (solid line), $\Delta=0.02$ (dashed blue line)
and $\Delta=0.1$ (dotted red line). Wide mode profiles convey a more adiabatic dynamics, causing a decrease in $P_{trans}$
and $S$. Here, $\Delta_z=15$, $\lambda_0=1$ and $k=\lambda_0/10$. }
\end{figure}

In this section we consider a more realistic situation. The
atoms perpendicularly traverse the field of a Fabry-Perot cavity, in
which the mode shape is given by a Gaussian \cite{haroche1}
\begin{equation}
\lambda(z)=\lambda_0\mathrm{e}^{-\frac{z^2}{L^2}}.
\end{equation}
The case of a Gaussian atom-field coupling in the mazer was
considered numerically in \cite{mazergauss}, and the tunneling
resonances where compared with the ones of the hyperbolic secant of
the previous section. We, however, use a completely different
numerical approach, namely a wave packet propagation procedure more
commonly used in molecular and chemical physics \cite{heller}. An
initial wave packet, which in our case is taken to be a minimal
uncertainty Gaussian
\begin{equation}
\Psi(z,t_i)=\left(\frac{1}{\pi\Delta_z^2}\right)\mathrm{e}^{-\frac{(z-z_0)^2}{4\Delta_z^2}}|e,0\rangle,
\end{equation}
is propagated in time under the corresponding Hamiltonian
(\ref{ham2}) using the {\it split operator method} \cite{split}.
Thus, the wave packet is obtained at any instant of time and
physical quantities are easily extracted from it. In particular,
the transmission probability is calculated as
\begin{equation}
P_{trans}=\lim_{t\rightarrow+\infty}\int_0^\infty\,dz\,|\Psi(z,t)|^2,
\end{equation}
where the time limit is taken such that the result has
asymptotically converged. An advantage of this approach is that it
automatically takes into account for motional quantum fluctuations
via the width $\Delta_z$ of the wave packet. Contrary to Reference
\cite{mazergauss}, we will be interested in the effect of having a
non-zero atom-field detuning $\Delta$ as well as the full system dynamics
during the scattering process.

\subsection{Tunneling}
As already pointed out, the meza mode profile is expected to give an
incorrect descripsion of realistic experimental situations. Firstly, it
does not have the actual shape of typical cavity mode profiles, and
secondly, it is a non-analytic function. Indeed, it is
known that discontinuities cause non-adiabatic behavior and cannot
correctly reproduce realistic cavity QED situations \cite{jonas4}.
More realistic results are instead expected to be obtained from the solvable hyperbolic secant model. However, this system is only solvable for
a zero detuning \cite{mazerdet}. Here we want to explore the scattering
characteristics in the case of a Gaussian mode profile and compare with earlier
results for the meza and hyperbolic secant.

In Figure \ref{fig6} we present the numerical results for the
transmission probability $P_{trans}(L)$ and the von Neumann entropy
$S(L)$. The solid lines give the zero detuning case where the
coupled equations can be decoupled into two scattering processes of
attractive and repulsive potentials respectively. The tunneling
resonances appears more often than for the hyperbolic secant
coupling of Figure \ref{fig5}, which was also pointed out in
\cite{mazergauss}. Another difference is the more rapid convergence
to the asymptotic values $P_{trans}(L)=1/2$ and $S(L)=1$ for large
$L$ of the hyperbolic secant profile compared to the Gaussian. This
is a consequence of the longer tails of the hyperbolic secant and
the smoother behaviour in general. This feature has been studied in
a semi-classical approach for cavity QED models in \cite{jonas4},
where it was found that the evolution is adiabatically more robust
for a hyperbolic secant than for a Gaussian mode profile. However,
it should be mentioned that the atomic c.o.m. motion was
replaced by an effective time-dependence and the atomic kinetic
energy exceeds the atom-field interaction energy in that work. The
dotted and dashed lines in Figure \ref{fig6} are the results obtained from having a non-zero detunings $\Delta$. Note that in both cases
$\Delta<\lambda_0$ and still a great effect of the non-zero detuning is found. In other words, the tunneling dynamics is very sensitive
to the parameter $\Delta$, and we especially note that this
detuning dependence is clearly stronger for a Gaussian mode profile
compared to a meza function \cite{mazerdet}. We discussed the
adiabatic and diabatic potentials in Section \ref{sec2}, where
$|\Delta|\rightarrow\infty$ gives the adiabatic limit. Figure
\ref{fig6} confirms our earlier conclusions that an increase in
adiabaticity depletes the tunneling effect; large $\Delta$ and $L$
render an adiabatic evolution and in this regime the tunneling
resonances are completely washed out.

\subsection{Dynamics}

The atom-field dynamics in the mazer was briefly considered in
\cite{tunneltime} in terms of the characteristic tunneling time for
a meza mode profile. Dynamics was also analyzed for an ultracold
atom traversing a cavity with a cosine mode profile consisting of
several periods \cite{jonas5}. It was found that due to the quasi
periodicity, an effective mass can be introduced in order to
understand the evolution. The same system was investigated further
in \cite{jonas6} studying the concept of a matter wave index of
refraction.

\begin{figure}[h]
\begin{center}
\includegraphics[width=12cm]{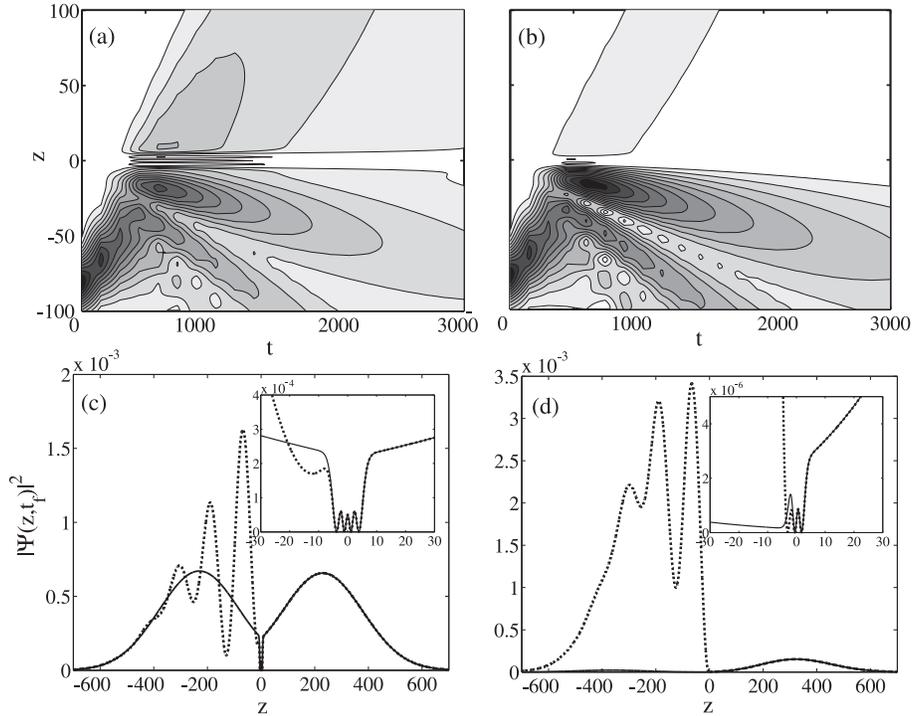}
\end{center}
\caption{\label{fig7} Time evolution of the probability distribution
$|\psi_e(z,t)|^2+|\psi_g(z,t)|^2$ (a) and (b), and the individual
amplitudes $|\psi_e(z,t=3000)|^2$ (dotted) and
$|\psi_g(z,t=3000)|^2$ (solid) in (c) and (d) for a Gaussian mode
profile. The left plots display the case where the parameters are
chosen to give maximum tunneling (corresponding to the third
resonance of Figure \ref{fig6}) while the right ones show the
results with parameters in between two resonances (second minima of
Figure \ref{fig6}). The insets zoom in on the wave packets in the
classically forbidden region. Other parameters are as in Figure
\ref{fig6}. }
\end{figure}

We now turn to consider the Gaussian mode profile and the
wave packet evolution while passing the interaction region, both
when the tunneling condition is fulfilled and when it is not. We use
the parameters of Figure \ref{fig6} and chose $L=2.1245$ or
$L=1.8108$ for the two situations respectively. These values of $L$
correspond to the third tunneling resonance and the second minima of
Figure \ref{fig6}. We restrict the analysis to the zero detuning
case. The results from the wave packet propagation are depicted in
Figure \ref{fig7}. First, in (a) and (b) we give the total wave
packet amplitude $|\Psi(z,t)|^2=|\psi_e(z,t)|^2+|\psi_g(z,t)|^2$,
while (c) and (d) show the constituent amplitudes
$|\psi_e(z,t=3000)|^2$ (dotted) and $|\psi_g(z,t=3000)|^2$ (solid)
at time $t=3000$. The insets give a close-up of the amplitudes
around the interaction region, $\lambda(z)\neq0$. The difference
between the two examples of being in or ouside the tunneling regime is evident from the plots. The typical decrease of the wave packet
inside the classically forbidden regime throughout the evolution is
also visible from the figure \cite{tunnel}. The symmetric
oscillations structure of the amplitude inside the forbidden region,
seen in the inset of (c), is a characteristic of the tunneling
resonance condition where the number of nodes depends on which
resonance peak one considers; the number of atomic de Broglie
wavelengths that fit the cavity length. Note that some oscillations
are also present in the forbidden regime also in (d), but one should
bare in mind the difference in amplitude (factor of 100) of the two
insets.

\subsection{Motional induced collapse-revivals}

It is well known from quantum optics, and especially from cavity
QED, that an uncertainty in the field state results in a collapse in
the atom-field Rabi oscillations \cite{cr,sz}. The various photon
number states that form the field state induce a series of different
Rabi oscillations causing the collapse. At later times, the various
terms comes back in phase, manifesting it selves in a revival.
Also for the mazer has the collapse-revival effect been analyzed
\cite{mazerrev}. In our model do we consider an initial vacuum and no
collapse-revival pattern can arise due to field fluctuations. We
will, however, show that the fluctuations in the atomic motional
state may form a collapse-revival structure.

\begin{figure}[h]
\begin{center}
\includegraphics[width=10cm]{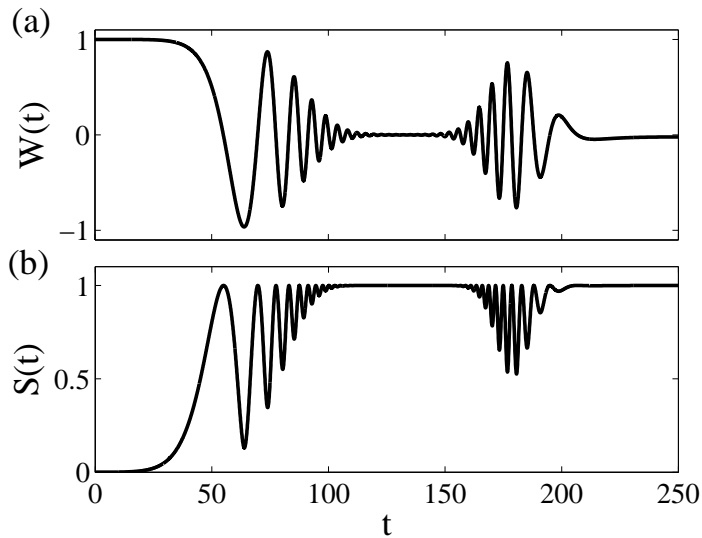}
\end{center}
\caption{\label{fig8} The atomic inversion $W(t)$ and the von
Neumann entropy $S(t)$ (for the Gaussian mode profile) as function
of time. Here $\Delta_z=10$, $\lambda_0=1$ $\Delta=0$ and $k_0=6$. }
\end{figure}

Assuming $k>\kappa_n$, but remaining in the regime where
$k\sim\kappa_n$, the whole wave packet will be transmitted through
the vacuum induced potential. From Figure \ref{fig4} (b) we have that, while traversing the
cavity field, the atom will Rabi oscillate with frequency
$\lambda(z)$. However, since the coupling $\lambda(z)$ is not
constant over the extent of the atomic wave packet, will the
oscillation frequency vary within the packet. Entering the
interaction region, the front of the wave packet sees a stronger
amplitude of the atom-field coupling than the back of it. Letting
$L/2$ and $\tau=L/2k_0$ denote characteristic length and time scales
respectively, and further assuming that the wave packet is to the
left of the origin $z=0$ and that $\Delta_z<L$. Then, typically if
$\big[\lambda(0)-\lambda(-\Delta_z)\big]\tau\sim\pi$ a collapse will
occur in the dynamical quantities, such as the {\it atomic
inversion} $W(t)=\langle\hat{\sigma}_z\rangle$ or the von Neumann
entropy. After passing $z=0$, the coupling decreases and as the
whole wave packet is to the right of the centre the situation will
be the opposite; the front of the packet oscillates with a smaller
frequency than the back. Thus, because of the symmetry of the
Gaussian, the collapse will be reversed resulting in a revival,
similar to {\it time-reversal echo processes} \cite{sz,mw}. The
numerical results shown in Figure \ref{fig8} for the inversion and
the von Neumann entropy confirm the above arguments of a
collapse-revival structer brought about due to momentum fluctuations. The imperfect revival is because wave packet spreading during the passage of the interaction region. Note that the revivals occurs due to the "time-reversal"
like dynamics and not because of a discreteness in quantum numbers,
as is the origin of regular revivals \cite{cr,robinet}.

\section{Conclusion}\label{sec5}
In this work we have investigated the mazer beyond earlier studies
by allowing for quantum fluctuations in the atomic motion and also
in the field statistics. Known analytical models, the meza or the
hyperbolic secant, have been extended to include such uncertainties
and their effects have been discussed. Typically, the
characteristics of various physical quantities, {\it e.g.}
entanglement or transmission probability, are smeared out by these
fluctuations. We have also shown that the fluctuations in general
increase the amount of atom-field entanglement.

The more physically interesting situation, in terms of experimental
realization, of a Gaussian mode profile was considered via a
numerical wave packet approach. This method naturally takes into
account for motional fluctuations and it is not limited to the
situation of zero atom-field detuning. A consequence of a non-zero
detuning is the fact that a decoupling of the internal two-level
structure is not straightforward, apart from the limiting cases of
adiabatic or diabatic evolution. We found that the system dynamics
is very sensitive to the value of the detuning, which opened up
questions about adiabaticity contra tunneling behaviour. Using our
numerical propagation procedure we also studied the dynamics of the
wave packet during the scattering against the vacuum potential.
Finally we demonstrated how motional fluctuation can induce novel
collapse-revival characteristics, different from the regular cavity
QED collapse-revivals originating from fluctuations in the field
statistics.

Experimentally, a difficulty with reaching the mazer regime is the
extremely long characteristic time scales. The atoms are cooled down
to the nK regime with typical c.o.m. velocity of mm/s \cite{mazer2}.
Using microwave cavities of mm-size \cite{haroche1} implies
operational times of a few seconds. This exceeds typical atomic and
cavity life-times which are of the order of ms. Another obstacle for
these long processes is gravity, which however may be circumvented
by using matter wave guides \cite{waveguid}. Atomic losses can be
minimized by using a two-photon {\it RAMAN} coupling, such that two
meta stable atomic states are dispersively coupled via an excited
state using one external laser field and the cavity
field \cite{bos}. This, however, introduces additional {\it Stark
shift} terms in the Hamiltonian. To overcome the problem of cavity
losses one could consider a pumped open cavity system. Another
possibility is to try to speed up the process, for instance by
considering a Bose-Einstein condensate coupled to the cavity
\cite{bec}, in which the effective atom-field coupling can be
increased by a factor 1000 compared to the single atom case
\cite{bec2}. In such a system, the intrinsic non-linearity leads to
novel phenomena like bistability. This is under current
investigation.

\ack

The author wishes to thank Profs. Stig Stenholm and Maciej
Lewenstein for interesting discussions

\end{document}